\documentstyle[preprint,aps,prb]{revtex}
\begin{document}
\input psfig
\draft
\title{\bf 
Magnetization of a Diffusive Ring: Beyond the Perturbation Theory}
\author{V. V. Afonin $^{(1)}$ and Yu. M. Galperin $^{(1,2)}$}
\address{$^{(1)}$ A. F. Ioffe Physico-Technical Institute, Russian Academy of
Sciences\\ $^{(2)}$ Department of Physics, University of Oslo, P. O. Box 1048
Blindern, 0316 Oslo, Norway}
\date{\today}
\maketitle
\begin{abstract}
Average persistent current over a set of diffusive metallic rings
with fixed number of electrons is considered. We study the the case
where the phase breaking time is much greater than an inverse average
inter-level distance.  In such a situation, many return events for an
electron have to be taken into account. As a result, one arrives at a
non-perturbative problem  for a fixed by an external magnetic field
Cooperon mode. This multi-Cooperon problem problem has been considered
previously by  Altland et al., Europhys. Lett., {\bf 2}, 155 (1992) and
in several 
following papers within the framework of supersymmetric approach. Such
an approach involves very tedious calculations which were performed
using computer algebra package.
Here we solve the problem in question with the help of replica trick. 
It is demonstrated that the replica trick in combination with a proper
analytical continuation in the replica space allows one to obtain the
result in much more explicit way.
\end{abstract}

\bigskip
\pacs{PACS numbers:72.10.Bg, 73.20.Dx}  
\narrowtext

\section{Introduction} 
\label{int} 

Magnetic properties of small conductors were extensively studied
during the last several years (see \cite{im,ch} and references
therein). It has been understood that the magnetic moment (and the 
associated persistent current) induced by an external magnetic flux
is a very specific manifestation of mesoscopic behavior. While
originally predicted to appear in clean one-dimensional metallic
rings \cite{kul:70}, most of the recent discussion about persistent
currents has focused on metallic rings containing impurities
\cite{levy:90}. Static magnetic properties of small rings and dots 
were studied by several
authors\cite{chan:91,Buttiker83,bm,Schmid,Oppen,agi,Ambegaokar}. 
An important step in the understanding of magnetization of mesoscopic
quantum rings was taking into account the difference between
canonical and grand canonical ensembles\cite{bm,agi,lb1,cgr,kg}. It
has been shown that the magnetization of isolated rings with fixed
number of particles is much larger than of the ensemble of rings kept
under fixed chemical potential. As a result, the main contribution to
the magnetic moment has been expressed in terms of the  fluctuation of
the number of particles at fixed chemical potential, 
$\langle (\delta N)^2\rangle$. The latter quantity has been
analyzed in\cite{agi,kg} under the condition 
$\hbar/\Delta \tau_\phi \gg 1$. Here $\Delta$ is an average
inter-level distance at the Fermi level, $\Delta^{-1}=\nu V$ ($\nu$ 
is the density of states at the Fermi level, while $V$ is the
volume). $\tau_\phi$ is the phase-breaking time. 

Let us  discuss the physical meaning of the parameter   
$\hbar/\Delta\tau_\phi$. As well known\cite{lk}, in the absence of
external magnetic field the quantum correction to the conductivity is
proportional to the classical probability $W$ for an electron with a
velocity $v$ and momentum $p$ to return to the vicinity of the
starting point (more exactly, into the volume of the order 
$v\, dt \, (\hbar/p)^{2}$ important for quantum interference). 
The probability $W$ is given by the expression\cite{lk}
\begin{equation}  
\label{prob}
W \propto \frac{v \hbar^2}{p^2}\int^{\tau_\phi} dt\,
P(\bbox{r},t)\left\vert_{\bbox{r}=0} \right.\, , 
\end{equation} 
where $ P(\bbox{r},t)$ is the probability density. Here we employ
the fact that in a diffusive regime it is a smooth function of
co-ordinates at the scale of mean free path $\ell$. To estimate
$P(0,t)$ we take into account that the electron diffusion is
restricted by finite volume of the sample. In such a case, we have  
\begin{equation} 
\label{1}
P(0,t) \propto \frac{1}{V}\sum_{n,
\bbox{n}_\perp} \exp
\left[-D\left(\frac{n^2}{R^2}+\frac{\bbox{n}_\perp^2}{d_\perp^2}
\right)t\right]\, .  
\end{equation} 
Here $D$ is the diffusion constant, $R$ is the radius of the ring,
while $d_\perp$ is its transverse dimension. 
The numbers $n,\bbox{n}_\perp$ have the meaning of quantum numbers
for longitudinal and transverse diffusive modes, respectively.  For a
thin ring, $d_\perp \ll R$, only  
$\bbox{n}_\perp = 0$ is important. One can see that at 
$D\tau_\phi/R^2 \gg 1$ one cannot replace the sum over discrete 
$n, \bbox{n}_\perp$ in (\ref{1}) by an integral. On the contrary, 
only $n=0$ is important, and $W \sim \tau_\phi \Delta/\hbar$. If 
this quantity is small, one can restrict himself with a single return
event. 

Now let us concentrate on the case of external magnetic field. In a 
magnetic field, the number $n$ in the expression (\ref{1}) has to be 
replaced by $n-\Phi/\Phi_0$, where $\Phi$ is the magnetic flux
embedded in the ring, while $\Phi_0=\pi \hbar c/e$.\cite{ash} It is
clear that the quantum contribution is maximal if $\Phi/\Phi_0$ is
close to some integer number $n_0$. If the difference 
${\tilde n} \equiv n_0 - \Phi/\Phi_0 = 0$ we arrive at the same
situation as in the absence of magnetic field -- only the mode with 
$n=n_0$ is important. One can expect that this property is also the
case at finite $|{\tilde n}| \ll 1$. Indeed, for $n \ne n_0$  
$$\delta W \propto \sum_{n \ne n_0}\frac{\Delta}{D{\tilde n}^2/R^2
+1/\tau_\phi} \sim \frac{\Delta R^2}{D} $$ (for the last estimate 
we have assumed $D\tau_\phi/R^2 \gg
1$). Consequently, if $\delta W
\ll 1$ one can neglect the contributions of all the modes with 
$n \ne n_0$ to the probability to return. However, the corresponding
contribution of the mode with $n =n_0$ is not small at $\Delta
\tau_\phi/\hbar \gtrsim 1$. Hence, we arrive at the problem of
calculation of the localization contribution in the case $$D/R^2 \gg
\Delta/\hbar \gtrsim 1/\tau_\phi \, .$$ In this region we can still use a
single-mode approximation, but the perturbation theory involving a
single return event fails. 

The problem in question was addressed by Altland et al.\cite{ss1,ss}
 (see also Refs.~\onlinecite{sz,pm,aie}). The authors
used the so-called  $Q$-Hamiltonian approach within the framework of
the supersymmetric method. 
An intrinsic feature of this method is that one has to cancel out specific
non-physical contributions. Therefore the supersymmetric
approach involves  tedious algebraic calculations.  Consequently, the
authors of \cite{ss1,ss} extensively used computer-algebra package.
As a result, intermediate equations
remain unpublished  because, as it was stated, the  computer printout
covers many pages.  

On the other hand, another approach - the so-called replica
method -- exists\cite{elk}. According to this approach,  one has to replace the  system
under consideration by $N$ systems  identical to the original one and
at the end tend $N \rightarrow 0$. Usually, after such a procedure one
arrives at relatively simple expressions. 
The limiting transition $N
\rightarrow 0$ (if done properly) automatically cancels out the
non-physical contributions that has to 
be done explicitly within the supersymmetric approach.

To take the full advantage of this property one needs a regular
procedure to calculate the limit $N \rightarrow 0$. The aim of the
present paper 
is to suggest a procedure  of analytical continuation of a
non-perturbative two-particle Green's function from integer $N$ to the
whole complex plane which includes the point $N=0$. Such a procedure
allows one to calculate the limit rather automatically excluding
necessity of direct cancelation of non-physical contributions. We
obtain an analytical non-perturbative expression for  persistent
current in a mesoscopic diffusive ring and compare it with the results
of \cite{ss1,ss}. 

The paper is organized as follows. In the
Section~\ref{be} basic equations for the fluctuation of the number
of particles, as well as for the persistent current are analyzed. The
effective action in the single-mode approximation is considered in
Section~\ref{ea}. In Section~\ref{acf} the particle number
auto-correlation function and persistent current are calculated in
the non-peturbative region. The results are summarized in Discussion.
In the following calculations we will put $\hbar =1$. Then $\hbar$
will be restored in estimates and final results. 

\section{Basic equations} 
\label{be}
According to\cite{agi}, the main contribution to the persistent
current $I$ can be expressed through the magnetic flux $\Phi$
embedded in the ring, as  
\begin{equation} 
\label{cdef} 
I = \frac{c\Delta}{2} \frac{\partial}{\partial \Phi} \langle
(\delta N)^2 \rangle_{\mu = \langle\mu\rangle}\, , 
\end{equation}   
where $\langle ( \delta N )^2 \rangle_{\mu = \langle \mu\rangle}$ is
the particle number auto-correlation function, calculated at a given
value of chemical potential. The latter can be expressed in terms of
single-electron Green's functions as\cite{as}  
\begin{equation} 
\label{n-cor}
\langle (\delta N)^2 \rangle = \int_{-\mu}^{0} d\epsilon_1 d
\epsilon_2 K(\epsilon_1,\epsilon_2) \, , 
\end{equation} 
where 
\begin{equation} 
\label{n-cor1} 
K(\epsilon_1,\epsilon_2)=\frac{1}{\pi^2}\int
d\bbox{r}_1d\bbox{r}_2 \, \{\langle \Im
G_{\epsilon_1}^R(\bbox{r}_1,\bbox{r}_1)\Im G_{\epsilon_2}^R
(\bbox{r}_2,\bbox{r}_2)\rangle 
- \langle
\Im G_{\epsilon_1}^R(\bbox{r}_1,\bbox{r}_1)\rangle \langle\Im
G_{\epsilon_2}^R
(\bbox{r}_2,\bbox{r}_2) \rangle\} \,. 
\end{equation}
Here $\langle \cdots\rangle$ means the usual impurity average. The
quantity (\ref{n-cor1}) has been calculated in Ref.~\onlinecite{agi}
in the limiting case $\Delta \tau_{\phi} \ll\hbar$. Our aim is to
go beyond this limiting case, namely to calculate the correlation
function for arbitrary $\Delta \tau_\phi/\hbar$, keeping $p_F \ell
\gg \hbar$. For this purpose we employ the approach by Efetov,
Larkin, and Khmelnitskii\cite{elk} with minor modifications. Namely,
we will use the so-called $Q$-Hamiltonian approach within the
framework of the replica trick. 
The confined expression for the correlation function $K(\omega)$
($\omega = \epsilon_1-\epsilon_2)$ can be written in the form (cf.
with Ref.~\onlinecite{elk}) 
\begin{equation}
\label{c-int1} 
K(\omega)= \left[\frac{\nu^2}{N^2\int DQ \exp(-
F)}\int d\bbox{r}_1 \, d \bbox{r}_2 \,
\int DQ e^{- F} \, \text{Tr} \,[\Lambda
Q(\bbox{r}_1)] \, \text{Tr} \, [\Lambda Q(\bbox{r}_2)]\right]_{N
\rightarrow 0}
\end{equation} 
where
\begin{equation}
 F=\frac{\pi \nu}{4}\int d\bbox{z}\, \text{Tr}
\,\left[D\left( \bbox{\nabla} Q +
\frac{ie}{c}\bbox{A}[Q,\Lambda]_-\right)^2
 +2 \left(i\omega-\frac{1}{\tau_\phi}\right) \Lambda Q
\right] \, . 
\end{equation}
 Here $\bbox{A}$ is the vector-potential, $[A,B]_{-} \equiv AB-BA$.
 Having in mind to take into account only an elastic
scattering by short-range non-magnetic impurities, we can specify  $Q$
as $2N\times2N$ Hermitian matrices, $Q^2=1$, 
$\text{Tr} \, Q=0$, $N$ is the number of replicas, while  
$$\Lambda= \left(\begin{array}{cc} \hat{1}&0\\0& -
\hat{1}\end{array}\right)\, ,$$
where $\hat 1$ is the $N \times N$ unit
matrix. The parameter $\tau_\phi^{-1}$ is introduced
phenomenologically. We assume that the phase breaking is due to
inelastic processes. Following\cite{elk}, we use the parameterization
$$Q=\Lambda \exp (W)\, ,\quad W =\left(\begin{array}{cc}0&B\\-B^+&0 \end{array} \right)\,,$$  
where $B$ is an arbitrary $N\times N$ matrix. 

\section{Effective action in single-mode approximation} 
\label{ea}
Consider a ring with the radius $R$ and the width $d_\perp\ll R$.
Consequently, one can take into account only the dependence of the
matrices $B$ on the
angular co-ordinate $\varphi$ . Expanding this
dependence into the discrete Fourier series, $B=\sum_n B_n
\exp(in\varphi)$, we introduce the mode number $n$. As it has been
explained in Section~\ref{int}, only one mode with $n=n_0$
corresponding to $\min (n-\Phi/\Phi_0)$ is important (this assumption
will be justified at the end of Section~\ref{vt}). Retaining only
this mode and assuming $\nabla_\varphi =
(1/R)\partial/\partial\varphi$ we get $W\nabla_\varphi W +
\nabla_\varphi W W =0$. Hence,  
$$
\nabla_\varphi Q 
=\frac{1}{R}\Lambda \frac{\partial}{\partial
\varphi}e^W=(\nabla_\varphi W) W^{-1}\sinh W 
=\frac{in_0}{R}\sinh W\, .
\nonumber  
$$
 Then,
 $\sinh^2 W=(1/2)(\cosh 2W -1)$ can be expanded as a series in 
$$W^{2k}=(-1)^k \left( \begin{array}{cc} \sqrt{B_{n_0} B_{n_0}^+}
& 0 \\
0 & \sqrt{B_{n_0}^+ B_{n_0}}
\end{array}\right)^{2k}\, .$$
The item $\text{Tr} \, (\Lambda Q)$ can be treated in a similar way.
As a result, we arrive at the following expression for $F$  
\begin{equation} 
\label{ac1} 
F
=\frac{\pi}{2 \Delta}\left[ \frac{D}{R^2}\left(n_0
-\frac{\Phi}{\Phi_0}\right)^2 \, \text{Tr}\, \sin^2
\left(\sqrt{B_{n_0}B_{n_0}^+}\right) 
+ 2\left(i \omega-\frac{1}{\tau_\phi}\right)\, \text{Tr}\,
\cos\left(\sqrt{B_{n_0}B_{n_0}^+}\right)\right] \, . 
\end{equation}  
Now we arrive at an important point. An arbitrary complex $N\times N$
matrix $B$ can be described by 2 Hermitian  $N\times N$ matrices. We 
defined them as 
\begin{equation} 
B =\rho \, \exp(i\varphi)\, , \quad B^+ =\exp(-i \varphi)\, \rho  \, .
\end{equation} 
The quantity $F$ is dependent only on the matrix $\rho$. On the other
hand, an arbitrary Hermitian matrix could be diagonalized, the
eigenvalues being {\em real}. One can immediately observe that the
integral over $\rho$ in the expression (\ref{c-int1}) with $F$ taken
from Eq.~(\ref{ac1}) for the correlation function diverges. This
divergence in fact does not occur, because the eigenvalues of $\rho$
must be defined within a finite interval. Indeed, one has to define
the variables $\rho$ in a way to obtain {\em one-to-one}
correspondence between $\rho$ and $Q$. On the other hand, one can
explicitly show that 
\begin {equation} 
\label{Q} 
Q=\Lambda e^W 
=\left( \begin{array}{cc} \cos \sqrt{B B^+} &
\frac{\sin \sqrt{B B^+}}{\sqrt{B B^+}} B \\B^+ \frac{\sin \sqrt{B
B^+}}{\sqrt{B B^+}}& -\cos \sqrt{B^+ B} \end{array}\right)
=\left( \begin{array}{cc} \cos \rho & \sin \rho
\, e^{i \varphi} \\ e^{-i \varphi}\sin \rho &-\cos \rho^T \, ,
\end{array}\right)  
\end{equation} 
where $\rho^T$ denotes the transposed matrix $\rho$. Here we have
employed the relationship 
\begin{equation} 
\label{rho-T} 
\rho^T = \sqrt{e^{-i \varphi}\rho^2 e^{i \varphi}} \, , 
\end{equation} 
which is a consequence of the symmetry properties of the initial
replica Hamiltonian (see Appendix~\ref{ham}). It is clear that the
matrix $Q$ is a periodic function of $\rho$, and one has to specify a
region at least not larger than 1 period to get one-to-one
correspondence. Moreover, to get proper analytical properties
(damping is the lower semi-plane of the $\omega$-variable) of the
action $F$ (\ref{ac1}) we have to define the integration limits as
$(-\pi/2,\pi/2)$. 
Finally, the action $F$ reads as   
\begin{equation} 
\label{ac2} 
F 
=\frac{\pi}{2 \Delta}\left[ \frac{D}{R^2}\left(n_0
-\frac{\Phi}{\Phi_0}\right)^2 \, \text{Tr}\, \sin^2 \rho 
+ 2\left(i \omega -
\frac{1}{\tau_\phi}\right)\, \text{Tr}\, \cos \rho\right] \, .  
\end{equation} 
Now let us transform the variables from $B,B^+$ to $\rho, u \equiv
\exp(i\varphi)$, the Jacobian being (see Appendix~\ref{det1a})  
\begin{equation} 
\label{det1} 
\frac{D(B,B^+)}{D(\rho, u)}= 2(\text{det} \  u^{-1} \rho)^N \, .
\end{equation}  
We observe that the variables $u$ can be integrated out and canceled
with the denominator in Eq.~(\ref{c-int1}).

\section{Particle number auto-correlation function} 
\label{acf} 
\subsection{Eigenvalue representation} 
\label{vt}

Now we come back to Eq.~(\ref{c-int1}). Taking into account that
$\text{Tr} \, \Lambda Q = 2 \, \text{Tr} \, \cos \rho $ we see that
the integrand is dependent only on eigenvalues of $\rho$. Hence, we
have to transform the variables to the eigenvalues and some other
ones which could be integrated out both in numerator and denominator.
This transform is outlined in Appendix~\ref{ev}. As a result, 
\begin{equation} \label{c-int2}K_N = \frac{(V \nu)^2\int_0^1 \{ d
\lambda\} \, \theta (\lambda^{(i)}) \, [\sum_{j=1}^N \cos(\pi
\lambda^{(j)}/2)]^2}{(N^2)\int_0^1 \{ d \lambda\}
\,\theta(\lambda^{(i)}) } \, , 
\end{equation}  
where $\{d \lambda\} \equiv\prod_{i=0}^{N-1}
d\lambda^{(i)}|\lambda^{(i)}|^{2i+N}$, while
\begin{equation}
\label{tx}
\theta (x) 
=\exp \left[ -\frac{\pi D}{2 \Delta
R^2}\left(n_0 -\frac{\Phi}{\Phi_0}\right)^2 \sin^2 \left(\frac{\pi
x}{2}\right) 
- \frac{\pi}{\Delta}\left(i ( \epsilon_1 -\epsilon_2)
-\frac{1}{\tau_\phi}\right) \cos \left(\frac{\pi x}{2}\right)
\right] \, . 
\end{equation} 
The expression (\ref{tx}) contains 3 dimensionless parameters,
\begin{equation} 
\gamma \equiv \frac{\pi \hbar}{ \Delta \tau_\phi}, \quad \Omega
\equiv \frac{\pi(\epsilon_1-\epsilon_2)}{\Delta}, \quad {\cal E}
\equiv \frac{\hbar \pi D {\tilde n}^2}{2R^2 \Delta}\, , 
\end{equation} 
where ${\tilde n} \equiv (n_0-\Phi/\Phi_0)$.
It is important to keep in mind the following.
If $\max{(\gamma, \Omega)} \gg 1$, only small $\lambda$ are
important. Hence, one arrives at the the result, obtained in the 
framework of perturbation theory\cite{agi,kg}. However, if both
$\gamma$ and $\Omega$ are small one has to sum multi-Cooperon
contributions that cannot be done in the framework of perturbation
theory. There is a substantial simplification in the case 
\begin{equation} 
\label{lc}
{\tilde n } \ll 1, \quad \text{but} \quad {\cal E}_c \equiv
\frac{\hbar \pi D}{2R^2 \Delta} \gg 1 \, .  
\end{equation} 
In this case, only one mode with $|n_0 - \Phi/\Phi_0| \ll 1$ is 
important, and it is the case where Eq.~(\ref{tx}) is valid.
Consequently, we consider the situation where the inequalities
(\ref{lc}) hold, but the quantities $\gamma$ and $\Omega$ can be
arbitrary. In fact, the mode $n_0$ must be considered by a
non-peturbative way, while the other modes can be treated within the
framework of the perturbation theory.

\subsection{Analytical continuation}

We are not able to calculate the expression (\ref{c-int2}) for an
arbitrary $N$ analytically. Instead, we will perform analytical
continuation of this expression to arbitrary $N$, and then calculate
its limit at $N \rightarrow 0$. 

\begin{figure} 

\centerline{\psfig{figure=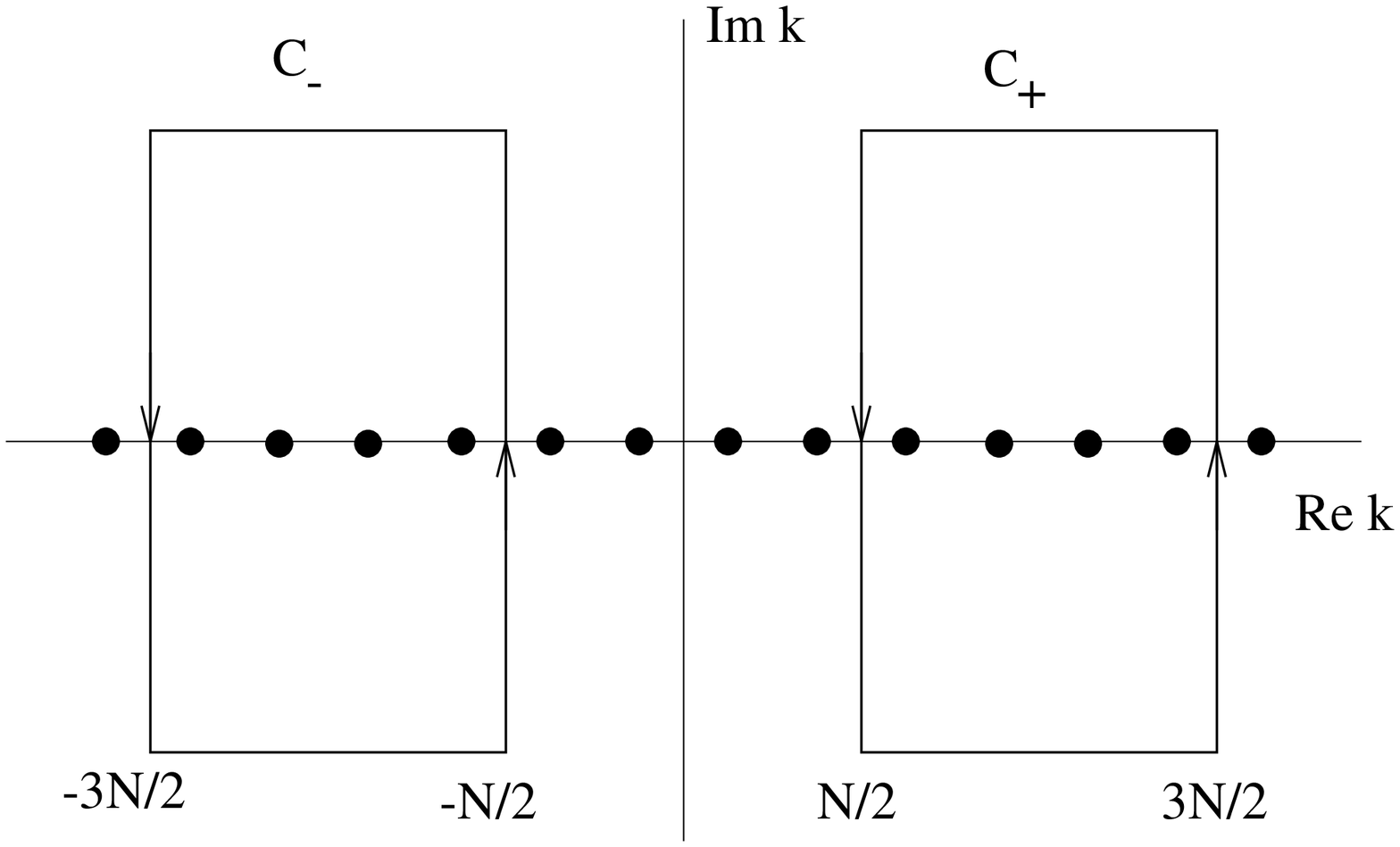,width=8cm}}  
\caption{The integration contours $C_\pm$ for $N=4$} 
\label{f1}
\end{figure}

\begin{figure}

\centerline{\psfig{figure=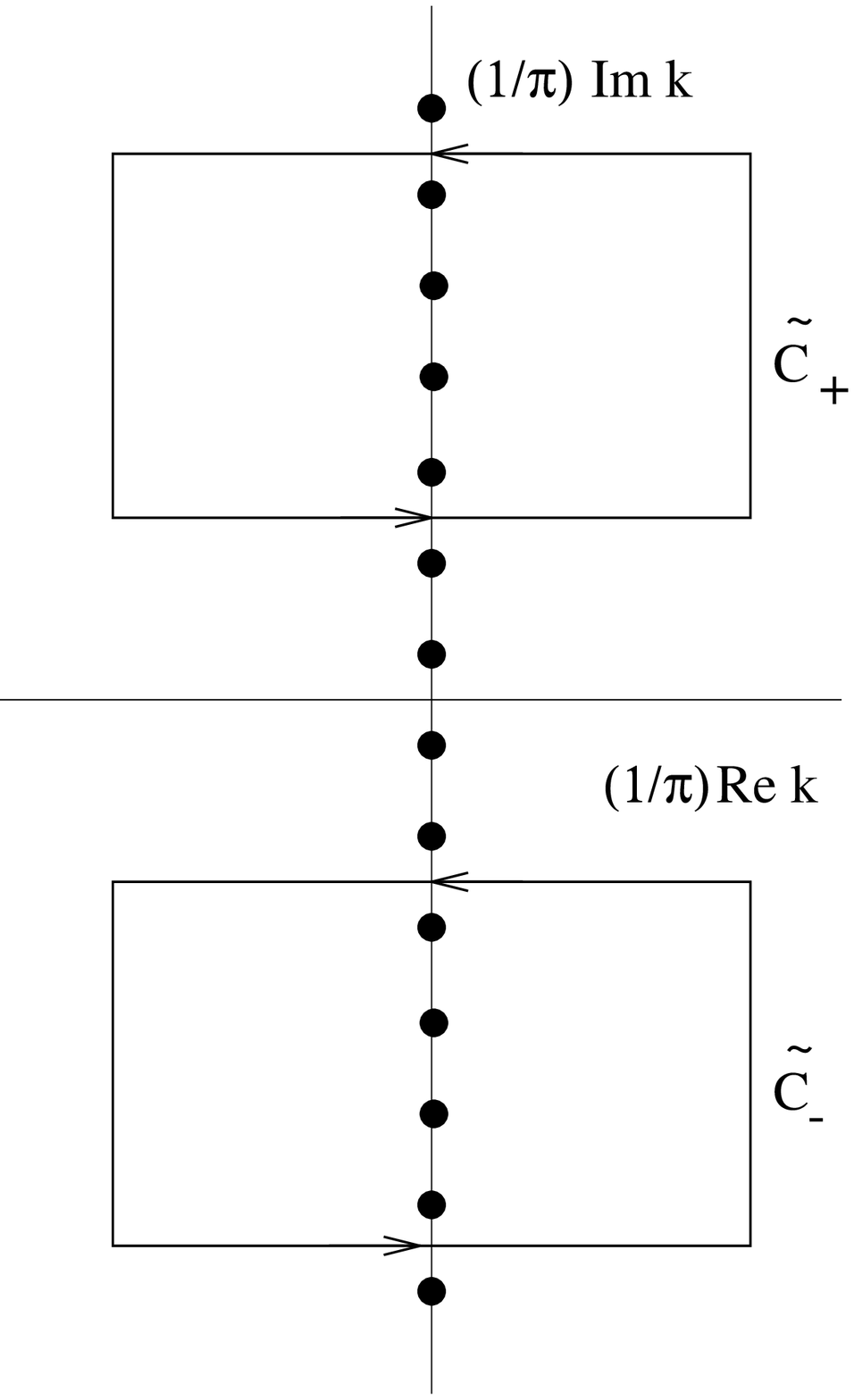,height=8cm}}

\caption{The integration contours ${\tilde C}_\pm$ for $N=4$}
\label{f2} 
\end{figure}

   Let us introduce the quantity
\begin{equation} 
\label{pf} 
Z_N = \prod_{k=0}^{N-1} \int_0^1 dx_k  x_k^{2k +N} \theta (x_k) \,,
\end{equation}  
where
\begin{equation} 
\theta (x) = \exp [-{\cal E}\sin^2(\pi x/2) -(i \Omega -
\gamma ) \cos(\pi x/2)] \, . 
\end{equation}
 It is convenient to define 
$$\zeta_N \equiv \ln Z_N \equiv \zeta^R + \zeta^A, $$ 
where 
\begin{eqnarray} 
\zeta^R &\equiv& \frac{1}{2}\sum_{k
=\frac{N+1}{2}}^{\frac{3N-1}{2}}\ln \int_0^1 dx \, x^{2k-1 +\delta}
\theta (x) \, ,
\nonumber \\
\zeta^A &\equiv& \frac{1}{2}\sum_{k
=-\frac{N+1}{2}}^{-\frac{3N-1}{2}}\ln \int_0^1 dx \, x^{-2k-1
+\delta} \theta (x) \, ,  
\end{eqnarray} 
$\delta $ is a small positive number which later will be put zero. We
introduce this parameter to keep the important integrals convergent
at the limit $N \rightarrow 0$. The first step is to express the sum
over $k$ in terms of contour integral over complex $k$. For this
purpose let us take into account that the derivative $\partial f[2\pi
i (k+1/2)]/\partial k$ [where $f(z)=(e^z +1)^{-1}$] has second-order
poles at integer numbers. Consequently, one can express
$\zeta_{R(A)}$ as  
\begin{eqnarray} 
\zeta_{R(A)}&=& \int_{C^\pm} dk \, \left(\frac{\partial f (2\pi
i k)}{\partial k} \right) F^\pm (k) \, , 
\\  F^{\pm} &=&\frac{1}{2}\int^k dk' \, \ln \left[ \int_0^1 dx\, x^{\pm
2k'-1 +\delta} \theta (x) \right]\, . 
\end{eqnarray}     
The contours $C^\pm$ are shown in Fig.~\ref{f1}. These expressions
are correct only if other singularities except of the poles of $f$
are unimportant. One can show that the function $F^+ (k)$ has
singularities only in the left-hand semi-plane of complex variable
$k$, while the function $F^-$ has singularities only in the
right-hand semi-plane of $k$. To prove this statement one has to
expand the function $ \theta (x)$ into a Taylor series. For the
following, it is convenient to rotate the $k$-plane by $\pi/2$ by
introducing a new variable $k_1 \equiv 2\pi i k$. The transformed
contours ${\tilde C}^\pm$ are shown in Fig.~\ref{f2}.

Making use of exponential convergence of the integral due to the
properties of $\partial f / \partial k$, we transform the contour
integrals to the integrals along the real axis. For simplicity let us
assume $N$ to be even. As a result, 
\begin{equation}
\zeta_{R(A)} 
= \mp \int_{- \infty}^{\infty} dk \,
\left(\frac{\partial f(k)}{\partial k} \right)  \left[F^{\pm}
\left( \frac{k \pm i\pi N}{2\pi i}\right)  
 -F^{\pm} \left( \frac{k \pm 3 i\pi N}{2\pi i}\right) \right] \, . 
\end{equation}
Now we are prepared to perform an analytical continuation over $N$.
We have to do it in a different way for the functions $F^\pm$ for the
reason to be discussed later. For this purpose we replace $i N$ by
$\pm N_0$ in the functions $F^\pm$, respectively. Here $N_0$ is a
real positive quantity which we are going to tend to zero later.
Finally, we have 
\begin{equation}
\zeta_{N_0} 
=\int_{- \infty}^{\infty} dk \, \left[
\left(\frac{\partial f(k- \pi N_0)}{\partial k} \right) -
\left(\frac{\partial f(k- 3 \pi N_0)}{\partial k} \right) \right] 
\left[F^{-} \left( \frac{k}{2\pi i}\right) -F^{+}
\left( \frac{k}{2\pi i}\right)\right] \, . 
\end{equation}   
As a result, the lowest-order term in the $N_0$-expansion of the
function $\zeta_{N_0}$ is $\propto N_0^2$. Finally, we get 
\begin{equation} 
\zeta 
= 2\pi i N_0^2 \int_{-\infty}^{\infty} dk \,
\left(\frac{\partial^2 f(k)}{\partial k^2} \right) 
\ln \left[\int_0^1 dx \, x^{-ik/\pi -1 + \delta}
\theta (x) \right] \, .  
\end{equation} 
The reason of splitting the function $\zeta_N$ into $\zeta^R$ and
$\zeta^A$ with the replacements $N \rightarrow \pm iN_0$ is as
follows. As $N$ tends to zero, the integration contour comes
infinitely close to the cut of logarithm functions which enter the
expressions for $F^\pm$. Such a situation is not the case for any
finite $N$, and it leads to a non-physical pinch which has to be
subtracted. Within the above mention procedure such a contribution is
pure imaginary while the one we are interested in is real. The
imaginary contributions to $F^+$ and $F^-$ have opposite signs. Thus
the non-physical contribution is automatically canceled out in the
sum $\zeta^R + \zeta^A$. We want to note that these terms are of the first order in
$N_0$. They have to vanishes, otherwise  the two-particles Green's function
would be divergent.
In fact, a similar  trick has been used by Matsubara to formulate the thermal
Green's function technique (see e.g., Ref.~\onlinecite{agd}). Let us compare our
analytical continuation of the function $\zeta$ to the  analytical
continuation of the two-particle Matsubara Green's function
$K(\Omega_m)$, where $\Omega_m$ is the external Matsubara frequency.
In both cases  one has to take two functions regular in the upper
(retarded) and the lower (advanced) semi-plane, respectively. Then one
combines the two above-mentioned functions into the one having a cut
in its complex plane. The physical reason of such a splitting into $R$
and $A$ parts is to cancel out  non-physical contributions. 
In the Matsubara case the non-physical contributions to $K(\Omega_m)$
arise in the point $\Omega_m \rightarrow 0$ and cancel out after 
the similar continuation $\Omega_m \rightarrow i\Omega$ of the sum
over $\Omega$.

\subsection{Persistent current}

Following\cite{agi}, we express the current according to
Eqs.~(\ref{cdef}),(\ref{n-cor}). On the other hand,
$$K(\epsilon_1,\epsilon_2) \propto
e^{-\zeta}\frac{\partial^2}{\partial \epsilon_1 \, \partial
\epsilon_2} e^{\zeta} \, .$$  Finally, we get 
\begin{equation}
I = - \frac{c \Delta}{2} \frac{\partial}{\partial \Phi} \lim_{N_0
\rightarrow 0} \frac {\zeta_{N_0}(\epsilon_1=\epsilon_2=0)}{N_0^2} =
- J_0 {\tilde n}{\cal G} \, ,  
\end{equation} 
where  
 $J_0 =  \hbar cD/R^2\Phi_0 =eD/\pi R^2$, while 
\begin{equation} 
\label{cur2} 
{\cal G}
=i\int_{-\infty}^{\infty} dk \, \frac{\partial^2 
f(k)}{\partial k^2}
\frac{\int_0^1 dx \, \sin^2 (\pi x/2) x^{-ik/\pi -1 +
\delta} \theta(x, \Omega=0)}{\int_0^1 dx \, x^{-ik/\pi -1 + \delta}
\theta(x, \Omega=0)} \, . 
\end{equation} 
One can check directly that at $$ \Delta \ll \frac{\hbar}{\tau_\phi}
\ll \frac{\hbar D}{R^2}, \quad |{\tilde n}| \ll 1 $$ the above
expressions lead to the expressions obtained in
Refs.~\onlinecite{agi,kg}. To show that one can calculate the
integrals with the help of steepest-descent approach to
get\cite{agi,kg}  
\begin{equation} 
\label{ag1} 
I= - J_1 {\tilde n}, \quad J_1 = \frac{e \Delta}{\pi^3}
\frac{D\tau_\phi}{R^2}\, .   
\end{equation} 
In the region 
$$\gamma \ll 1, \quad {\cal E}_c \gg {\cal E} \gg 1 $$ 
one can also develop a perturbation theory. Indeed, only small $x$ in
the integrals in Eq.~(\ref{cur2}) are important. The physical reason
of this fact is the magnetic-field-induced phase-breaking. In this
region, we arrive at the result 
\begin{equation} 
\label{cur3} 
{\cal G}=\frac{1}{\pi^2 {\tilde
n}^2}\, \frac{R^2 \Delta}{\hbar D}\, , \quad I=-
\frac{1}{\pi^3  {\tilde n}}\, \frac{e\Delta}{\hbar}\, . 
\end{equation}    
This result agrees with the asymptotic result of Ref.~\onlinecite{kg}
for $\gamma \gg 1, \ \sqrt{R^2 /D\tau_\phi} \ll {\tilde n} \ll 1$.
Note that the result (\ref{cur3}) obtained for $\gamma \ll 1$ is
valid in the region $$ \sqrt{R^2 \Delta/\hbar D} \ll {\tilde n} \ll
1\, . $$ 
For the case $\gamma \ll 1 , \ {\cal E} \ll 1$, where the
perturbation theory is not applicable, one can put $\theta(x,
\Omega=0)=1$. As a result, we get ${\cal G} =0.21$, the current being
\begin{equation} 
\label{r1} 
I= - 0.21 \frac{e D}{\pi^3 R^2}   {\tilde n} \, . 
\end{equation} 
We observe a maximum at ${\tilde n} \sim \sqrt{R^2 \Delta/\hbar D}$,
the maximal current being 
\begin{equation} 
I_{\max} \sim e \sqrt{\Delta D/\hbar R^2} \, .  
\end{equation} 
Expressions (\ref{cur3}) and (\ref{r1}) are fully consistent with the
curve calculated in\cite{ss1,ss} with a help of computer algebraic
package. Let us 
discuss the dependence of the maximal current on $ \gamma
\approx \hbar/\Delta \tau_\phi $. At $\gamma \gg 1$ the perturbation 
theory\cite{agi,kg} predicts the maximum of the current at ${\tilde
n} \sim \sqrt{R^2/D \tau_\phi}$, the maximal value being 
\begin{equation}  
I_{\max} \sim e\frac {\Delta}{\hbar}\sqrt{\frac{ \tau_\phi D}{R^2}}
\, .  
\end{equation} 
Consequently $I_{\max} \propto \gamma^{-1/2}$ at $\gamma \gg 1$, and
it $\gamma$-independent at $\gamma \ll 1$.  In this region, the
persistent current can be estimated also as $$J_{\max} \sim
\frac{ev_F}{R} \sqrt{\frac{\Delta \tau_{\text{el}}}{\hbar}}$$ where
$\tau_{\text{el}}$ is the elastic relaxation time. The quantity
$\Delta\tau_{\text{el}}/\hbar$ for a typical metal can be estimated
as $(\ell/R)(a^2/A)$ where $\ell$ is the mean free path, $a$ is a
typical inter-atomic distance, while $A$ is the cross-section of the
ring. Eq. (\ref{r1}) shows that at $\Delta \tau_\phi/\hbar \gg 1$ the
phase-breaking time $\tau_\phi$ does not enter the expression for the
persistent current. Coming back to Eq.~(\ref{prob}) we have to
conclude that at $t \gtrsim \hbar/\Delta$ the electronic wave packet
does not smear in space. That means that the $n_0$-mode of the
Cooperon is localized in some sense, the localization length being of
the order $\sqrt{\hbar D/\Delta}$. Of course, it does not mean
complete localization because other modes are still under weak
localization conditions. 

The range of parameters where the theory above is applicable and
leads to non-trivial results can be expressed as $$1 \ll R/\ell \ll
K, \ (1/K)(\tau_\phi/\tau_{\text{el}}) \, , $$ where $K \sim
(pd_\perp/ \hbar)^2$ is the number of transverse channels. The left
inequality is the criterion for a diffusion motion, the first right
inequality is just the Thouless criterion ${\cal E}_c \gg 1$, while
the last right one is the condition $\Delta \tau_\phi /\hbar \gg 1$.
One can see that one needs low temperatures to meet the inequality 
$\tau_\phi /\tau_{\text{el}} \gg K \gg 1$, as well as samples of very
small size. As far as we know, no previous experiments satisfy this
set of conditions. 

\section{Discussion} 

As one can see from the preceding sections, the results of  
the replica procedure being complicated for arbitrary  integer N are
rather simple in the limit $N \rightarrow 0$. 
In this limit, the non-physical contributions are  canceled out
automatically, while within the sypersymmetic method that has been
done this explicitly.  
An important feature that leads to such a simplification is the
employed procedure of analytical continuation which
has been done before  direct calculations. Namely, one has to 
{\em two} functions analytical in the upper (lower) semi-plane of the
complex plane of $N$, respectively. The proper analytical continuation
is a combination of these functions. Consequently, in has a cut at
$\Im N =0$. The procedure above  allows one to cancel out
automatically the non-physical pinch in the two-particles Green's
function, which otherwise would exist  at  $N_0 = 0$. We believe that
such a construction is important in general for the calculations
involving the replica trick.  
In such a way we reproduce analytically and rather simply the results
obtained in\cite{ss1,ss} by a  computer algebraic package.    

\acknowledgements  
We are thankful to V.~L.~Gurevich, V.~Yu.~Petrov, and A.~D.~Mirlin for
valuable comments. One of the author (V.V.A.) is grateful to the
Research  Council of Norway for a financial support within the
Cultural Exchange Program (KAS).   

\appendix 

\section{Effective action -- derivation}  
\label{ham} 
Here we re-derive the expression (\ref{c-int1}) following\cite{elk}
to make clear important symmetry properties. Following\cite{elk} we
use the replica trick and introduce a field operators as $$
\bbox{\Psi} =\{\psi_1,\ldots,\psi_N,\psi^+_1,\ldots,\psi_N^+\}, \quad
\bbox{\Psi}^+ = \left(\begin{array}{c} \psi_1^+ \\ \ldots \\ \psi_N^+
\\ - \psi_1 \\ \ldots \\ -\psi_N\end{array} \right)\, . 
$$ Here $\psi_i \psi_i^+ + \psi_i^+ \psi_i =0$. The action can be
written as 
\begin{eqnarray} 
F&=&i\int (d \bbox{r}) \, \bbox{\Psi}^+(\bbox{r})({\hat E}-{\hat{\cal
H}})\bbox{\Psi}^+(\bbox{r}) \, ,  
\nonumber 
\\ {\hat E}&=&E {\hat I}\, , \nonumber \\ {\hat{\cal H}}& =
&[H_0+U_{\text{el}}(\bbox{r})]{\hat I} - \left( \frac{\omega}{2}+ i
\delta \right) \Lambda \, .  
\end{eqnarray} 
Here $H_0$ is the free-electron Hamiltonian, $\hat I$ is the $2N
\times 2N$ unit matrix, while $U_{\text{el}}(\bbox{r}) = U_0
\sum_i^{M} \delta (\bbox{r}-\bbox{r}_i)$, $M$ being the total number
of impurities. The first $N$ rows of $\hat{\cal H}$ describe an
evolution of the retarded Green's functions, while the last $N$ rows
describe an evolution of the advanced one. The following step is
averaging over the positions of impurities. We have 
\begin{equation} 
\Sigma 
\equiv \prod_{i=1}^M \int \frac{d \bbox{r}_i}{V}\, \exp
\left[i U_0 \sum_{i=1}^M
\bbox{\Psi}^+(\bbox{r}_i)\bbox{\Psi}(\bbox{r}_i)\right] 
= \left[\int \frac{d \bbox{r}}{V}\prod_{f=1}^{2N} \left(1 + i
U_0 \psi_f^+ (\bbox{r})\psi_f (\bbox{r}) \right)\right]^M 
\end{equation} 
Here we have taken into account that only linear terms in $\psi_f^+$
and $\psi_f$ can enter the continual integral for the correlation
function (Grassman algebra). Because of the same reason, one has to
allow only for the terms with different $f$ while calculating the
product. For a weak scattering and in the thermodynamic limit $M,V
\rightarrow \infty, \ , M/V =\text{const}$, 
\begin{eqnarray}  
\Sigma &\approx& \exp (\delta \mu + i \Gamma ) \, , 
\nonumber \\ 
\delta \mu &=& \frac{MU_0}{V}\int d \bbox{r} \,
\bbox{\Psi}^+(\bbox{r})\bbox{\Psi}(\bbox{r})\, , 
\nonumber \\
 \Gamma &=& \frac{g_0^2}{2}\int d \bbox{r}\sum_{f \ne
g}
\psi^+_f(\bbox{r})\psi_f(\bbox{r})\psi^+_g(\bbox{r})\psi_g(\bbox{r})
\, .  
\end{eqnarray}  
Here $\delta \mu$ is a shift in chemical potential, while $g_0^2 = 2
MU_0^2/V$ ($g_0$ is the coupling constant). In a same way as it has
been done in Ref.~\onlinecite{elk}, we introduce an auxiliary scalar
field represented by Hermitian matrices $Q$. As a result, the
effective $\psi^4$ interaction can be decoupled as \widetext    
\begin{eqnarray}  
\label{id1} 
&&\exp \left[ - \frac{g_0^2}{2}\int d \bbox{r}\sum_{f \ne g}
\psi^+_f(\bbox{r})\psi_f(\bbox{r})\psi^+_g(\bbox{r})\psi_g(\bbox{r})
\right] 
\nonumber  
\\ &&
= \frac{ \int {\cal D}Q \, \exp \left[- \text{Tr} \, \int d
\bbox{r} \left(\frac{\pi \nu}{4 \tau_{\text{el}} } Q^2 (\bbox{r}) -
\frac{1}{2 \tau_{\text{el}}}
\bbox{\Psi}^+(\bbox{r})Q\bbox{\Psi}(\bbox{r})\right) \right] }{ \int
{\cal D}Q \, \exp \left[- \text{Tr} \, \int d \bbox{r} \frac{\pi
\nu}{4 \tau_{\text{el}}} Q^2 (\bbox{r}) \right] } \, . 
\end{eqnarray} 
\narrowtext 
Here we have used the definition $2\pi \nu g_0^2 \tau_{\text{el}}
=1$. This expression is just the same as Eq.~(17) from\cite{elk}.  To
analyze symmetry properties of the impurity-averaged Hamiltonian let
us take into account that the initial Hamiltonian possesses the
property $\left.{\cal H}_{ij}\right|_{i,j \le N} ={\cal
H}_{i+N,j+N}^*$. Such a property has to be kept after impurity
averaging and introducing the field $Q$. In terms of $Q$ it reads as
$iQ_{ij}= -iQ_{i+N,j+N}^*$. Taking into account Eq.~(\ref{Q}) we
arrive at the relationship (\ref{rho-T}). The following steps are
exactly the same as in\cite{elk}.  

\section{Calculation of the Jacobian}  
\label{det1a} 
Let us arrange the columns of the $2N^2 \times 2N^2$ matrix $\partial
(B,B^+)/\partial (\rho, u)$ as
$$\{B_{11},\ldots,B_{N1};B_{12},\ldots,B_{NN};B_{11}^+,\ldots,B_{NN}^+\}$$
and
the rows as 
\begin{eqnarray} 
&&\{\rho_{11},\ldots,\rho_{N1};\rho_{12},\ldots,\rho_{N2},\ldots,\rho_{NN};
\nonumber 
\\&& \quad u_{11},\ldots,u_{N1};
u_{12},\ldots,u_{N2},\ldots,u_{NN}\}. 
\nonumber 
\end{eqnarray} 
Taking into account the matrix identities 
\begin{eqnarray} 
dB&=& d\rho \, u + \rho \, du , 
\nonumber 
\\  dB^+&=& u^{-1}\, d\rho - u^{-1}\, du \, u^{-1}\rho 
\end{eqnarray} 
we express $\partial (B,B^+)/\partial (\rho, u)$ as  
\begin{equation} 
\frac{\partial (B,B^+)}{\partial (\rho, u)} = \left(
\begin{array}{cc} {\hat A}_{11} &{\hat A}_{12} \\ {\hat A}_{21}
&{\hat A}_{22} 
 \end{array} \right) 
\end{equation} 
where ${\hat A}_{ik}$ are $N^2 \times N^2$ matrices. One can show
that 
\begin{eqnarray} 
{\hat A}_{11} &=& u \times {\hat 1}, \quad {\hat A}_{22}=
-(u^{-1}\rho) \times u^{-1} \, , 
\nonumber  
\\ {\hat A}_{21}&=& \underbrace{\rho \otimes \rho \otimes \ldots
\rho}_{N}\, , 
\nonumber  
\\ {\hat A}_{12}&=& \underbrace{u^{-1} \otimes u^{-1} \otimes \ldots
u^{-1}}_{N}\, . 
\end{eqnarray}  
Here $\times$ means the Kronecker product, while $\otimes$ means
direct product\cite{lv}. Making use of the identity\cite{lv}
$$\text{det} \, (A \times B) \equiv (\text{det} \, A)^p \,
(\text{det} \, B)^q $$ (where $q,p$ are the ranges of the matrices
$A$ and $B$, respectively), as well as the Laplace expansion of
determinant\cite{lv} we arrive at Eq.~(\ref{det1}). 

\section{Variable transformation}
\label{ev} 
Consider the set of variables which includes eigenstates
$\lambda_{(b)}$ and $N$ eigenvectors $\bbox{X}^{(b)}$. One can see
from the definition $\rho_{ik} X^{(b)}_k=\lambda^{(b)}
\delta_{ik}X_i^{(b)}$ that any vector of the type $e^{i \chi^{(b)}}
\bbox{X}^{(b)}$ (where $\chi^{(b) }$ is an arbitrary phase) satisfies
the equation with the same $\lambda^{(b)}$ and $\rho_{ik}$.
Consequently, one must exclude $N$ extra variables $\chi^{(b)}$. For
this purpose we require the diagonal elements $X^{(i)}_i$ to be {\em
real}. Consequently, the matrix $X^{(i)}_k$ can be constructed
according to the following procedure. The first column, $X^{(1)}_i$,
contains $N-1$ variables $X^{(1)}_i, \, i \ne 1$, while the last
(real) one, $X^{(1)}_1$ is calculated from the requirement of
normalization. In the next column, $X^{(2)}_i$, the last $N-2$
variables are chosen are independent. The element $X_1^{(2)}$ is
determined by the orthogonality of the vectors $\bbox{X}^{(2)}$ and
$\bbox{X}^{(1)}$, while the last element, $X_2^{(2)}$ is determined
by the normalization of $|\bbox{X}^{(2)}|$. The following elements
are determined by continuation of this procedure. Note, that all the
off-diagonal elements are complex ones, so one can consider real
($U^{(i)}_k$) and imaginary (($V^{(i)}_k$) parts. In this way we
present $N^2$ independent elements of the matrix $\rho$ through $N$
eigenvalues $\lambda^{(i)}$, and $N^2 - N$ independent variables
$U^{(i)}_k$ and $V^{(i)}_k$. 
{}From the definition, $\rho_{ij}=\sum_k
X_i^{(k)}\lambda^{(k)}X_j^{(k)*}$, one can express $\rho$ through
$\{U,V,\lambda\}$ as 
\begin{eqnarray} 
\label{tf} 
\frac{\partial \rho_{ij}}{\partial
U_i^{(k)}}&=&\lambda^{(k)}[U_j^{(k)}(1+\delta_{ij})-iV_j^{(k)}
(1-\delta_{ij})]\, , \nonumber \\  
\frac{\partial \rho_{ij}}{\partial
V_i^{(k)}}&=&\lambda^{(k)}[iU_j^{(k)}(1-\delta_{ij})+V_j^{(k)}
(1+\delta_{ij})]\, , 
\nonumber \\  
\frac{\partial \rho_{ij}}{\partial \lambda^{(k)}}&=&
U_i^{(k)}U_j^{(k)}+V_i^{(k)}V_j^{(k)} 
+i(V_i^{(k)}U_j^{(k)}-U_i^{(k)}V_j^{(k)}) \, . 
\end{eqnarray}   
Note that the above formulas do not contain summation over repeated
superscripts. To calculate the Jacobian we arrange the corresponding
$N^2\times N^2$ transformation matrix in the following way. The
columns are labeled by $N^2$ ``old'' variables  
$$\{\rho_{i1},\rho_{i2}, \ldots, \rho_{iN}\}.$$ 
The rows are labeled by $N^2$ ``new'' variables \widetext
$$\underbrace{\{U^{(1)}_{k > 1}\}}_{N-1},\underbrace{\{U^{(2)}_{k >
2}\}}_{N-2}, \ldots, \underbrace{\{U^{(N-1)}_N\}}_{1};
\underbrace{\{V^{(1)}_{k > 1}\}}_{N-1},\underbrace{\{V^{(2)}_{k >
2}\}}_{N-2}, \ldots, \underbrace{\{V^{(N-1)}_N\}}_{1};
\underbrace{\{\lambda^{(k)}\}}_N \, . $$ \narrowtext Consequently, as
follows from Eq.~(\ref{tf}), the first $(N-1)$ rows contain the
common factor $\lambda^{(1)}$ times quantities which depend only on
$\{U,V\}$. The next $(N-2)$ lines contain the factor $\lambda^{(2)}$,
and so on. The last $N$ lines are $\{\lambda\}$-independent. As a
result, the Jacobian can be expressed as $\prod_{i=1}^N
\left[\lambda^{(i)}\right]^{2(N-i)}\times$(some function of
$\{U,V\}$). This expression has to be multiplied by $(\text{det}\,
\rho)^N =\left(\prod_{i=1}^N \lambda^{(i)}\right)^N$, and we arrive
at Eq.~(\ref{c-int2}).

\newpage 

\end{document}